\documentclass[runningheads]{llncs}

\usepackage{graphicx}
\usepackage{tikz}
\usepackage{pgfplots}
\usepackage{tabularx}

\begin{document}

\title{Automatic Classification of Error Types in Solutions to Programming Assignments at Online Learning Platform}

\titlerunning{Automatic Classification of Error Types in Programming Assignments}

\author{
Artyom Lobanov\inst{1,2} \and
Timofey	Bryksin\inst{1,3} \and
Alexey Shpilman\inst{1,2} 
}

\institute{
JetBrains Research, Saint Petersburg, Russia\\
\email{alexey@shpilman.com}
\and
Higher School of Economics, Saint Petersburg, Russia \\
\email{avlobanov@edu.hse.ru}
\and
Saint Petersburg State University, Saint Petersburg, Russia \\
\email{t.bryksin@spbu.ru}
}

\authorrunning{A. Lobanov et al.}

\maketitle       

\begin{abstract}
Online programming courses are becoming more and more popular, but they still have significant drawbacks when compared to the traditional education system, e.g., the lack of feedback. In this study, we apply machine learning methods to improve the feedback of automated verification systems for programming assignments. We propose an approach that provides an insight on how to fix the code for a given incorrect submission. To achieve this, we detect frequent error types by clustering previously submitted incorrect solutions, label these clusters and use this labeled dataset to identify the type of an error in a new submission. We examine and compare several approaches to the detection of frequent error types and to the assignment of clusters to new submissions. The proposed method is evaluated on a dataset provided by a popular online learning platform.
\end{abstract}


\section{Introduction}
Recently more and more people get additional education through massive online open courses (MOOC), including programming courses. They are very convenient for students, but you get less feedback on what you are doing wrong since the solutions are usually checked using an automated verification system. In our study, we propose an automatic data-driven method for error type classification that can be used to provide hints for students, rather than just inform them on whether or not their submission has passed all the necessary tests. 

\begin{figure}[h]
  \centering
  \includegraphics[width=0.7\textwidth]{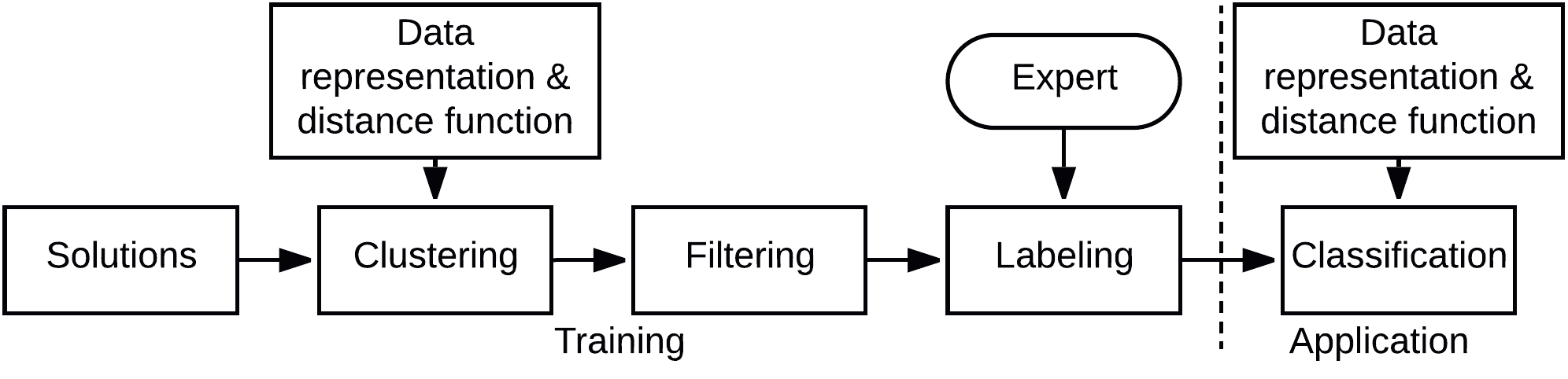}
  \caption{General pipeline of the proposed approach.}
  \label{fig:pipeline}
\end{figure}

The main idea of the proposed approach is to automatically identify and recognize the most common errors through identifying \textbf{edit scripts} and analyze these edits through clustering. We use expert evaluation to assign error types to clusters.  The general pipeline for the process can be seen in Figure~\ref{fig:pipeline}.

\section{Related work}
 Several papers have investigated topics related to our task, namely representing code changes in a vector form, their clustering, and classification.

Falleri et al.~\cite{GumTree} introduced an approach and a tool they called GumTree to generate \textbf{edit scripts}: sequences of atomic abstract syntax tree (AST) modifications that turn a source tree into a target tree. Today, GumTree is considered to be a state-of-the-art tool for generating edit scripts and is widely used in different research tasks.

The task of clustering source code changes based on edit scripts was studied in~\cite{ChangesClustering}. One of the considered approaches is similar to the one used in our work: edit scripts were generated using ChangeDistiller~\cite{ChangeDistiller} (a predecessor of GumTree) and a similarity-based clustering algorithm was applied to them. 

Another related task is the classification of code changes. In~\cite{ManualClassification}, the authors try to detect code changes that are likely to be specific types of refactorings. They define heuristic rules to define each refactoring type. This kind of approach could not be applied to our case because the number of possible errors and their types are not known beforehand.
In~\cite{CleanOrBuggy}, the authors used an SVM classifier and features such as commit's metadata, complexity metrics and bag-of-words of the changed code to identify commits that are likely to introduce new bugs. The authors treated code as text using bag-of-words models, whereas working with an AST usually gives more useful information. The somewhat similar idea was implemented in~\cite{StatisticsBagPrediction}, where features of AST changes were used for bug prediction. Their experiments also confirmed that using AST features rather than text-based ones yields better results.

One recent study~\cite{EditRepresent} provides an alternative way to represent edit scripts. The authors employ a deep learning approach to generate a vector of features (embedding) for the edit scripts.

\section{Overview of the approach}

\subsection{Dataset}\label{dataset}
The dataset is provided by Stepik\footnote{Stepik MOOC platform: \url{http://stepik.org/}} and consists of submitted solutions in Java with their metadata, including verification result. We follow the assumption that between the first correct solution and the previous (incorrect) one the user fixed a mistake, therefore changes between these versions contain information about a correctable error.

The dataset consists of \textit{\{incorrect submission, correct submission\}} pairs for 2 tasks: 1472 pairs for the problem A (a Java Stream API problem) and 8294 pairs for the problem B (checking double values for equality). The dataset was divided into train/validation/test subsets: 1176/148/148 pairs for the problem A and 6588/200/200 pairs for the problem B respectively.

\subsection{The pipeline}\label{pipeline}

The pipeline is divided into two stages: training and application. At the training stage, we try to find the most common errors in the incorrect solutions database. To achieve this goal, we cluster edit scripts for incorrect solutions. Edit scripts for the same error are expected to fall into the same cluster. At the next stage, labeled clusters are used to create a classifier. This classifier outputs a type for a new error if it falls into one of the identified clusters, and labels this error as ``unknown'' otherwise. 

To generate edit scripts we used the GumTree library. Since users can fix errors differently, at the clustering step we calculate edit scripts between an incorrect solution and all the correct solutions in the dataset and select the shortest edit scripts. The same procedure is used when we try to identify edit scripts for new incorrect solutions at the classification step.

In this paper, we used classification and clustering algorithms that require only a distance function defined between data points. We considered the following distance metrics for edit scripts: 
\begin{enumerate}
\item several modifications of the Jaccard similarity coefficient depending on the definition of equality of atomic changes in edit scripts;
\item cosine similarity for the bag-of-words model;
\item cosine similarity for the autoencoder embeddings of the edit scripts.
\end{enumerate}

We use Hierarchical Agglomerative Clustering (HAC)~\cite{DataCLustering} to cluster all solutions edit scripts using one of the distance functions described above. After the clustering is complete, clusters smaller than a certain threshold are removed and others are presented to experts, who label them according to the error type. 

When classifying a new incorrect solution, we find the nearest correct one and classify the obtained edit script. The easiest way to choose the right cluster is to find the nearest one. As an alternative, we used the k-nearest neighbors (kNN)~\cite{kNN} method with weighted voting. Since the new object may not belong to any cluster, we provide a user with a hint only if we are sure of the classification accuracy.

\section{Evaluation}
To evaluate various clustering and classification algorithms and their parameters we used the area under the precision-recall curve (PR-AUC) since classification should be tuned according to the particular goals. 

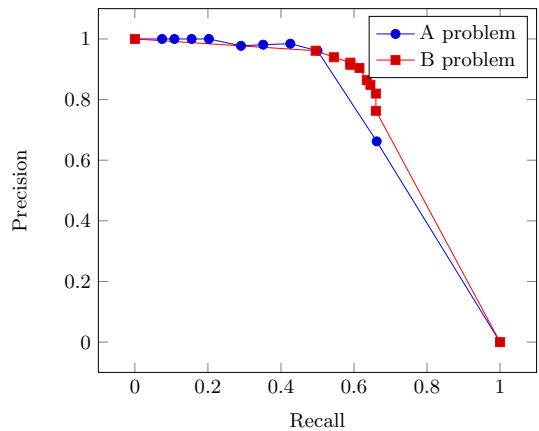
\begin{figure}[h]
  \centering
\begin{tikzpicture}[scale=0.85]
\begin{axis}[
ylabel=Precision,
xlabel=Recall] 
\addplot coordinates {
(0, 1)
(0.07432432432432433, 1.0)
(0.10810810810810811, 1.0)
(0.1554054054054054, 1.0)
(0.20270270270270271, 1.0)
(0.2905405405405405, 0.9772727272727273)
(0.35135135135135137, 0.9811320754716981)
(0.42567567567567566, 0.984375)
(0.5, 0.961038961038961)
(0.6621621621621622, 0.6621621621621622)
(1, 0)};
\addlegendentry{A problem}

\addplot coordinates {
(0, 1)
(0.495, 0.9611650485436893)
(0.545, 0.9396551724137931)
(0.59, 0.921875)
(0.59, 0.9147286821705426)
(0.615, 0.9044117647058824)
(0.635, 0.8639455782312925)
(0.645, 0.8486842105263158)
(0.66, 0.8198757763975155)
(0.66, 0.7630057803468208)
(1, 0)};
\addlegendentry{B problem}
\end{axis}
\end{tikzpicture}

  \caption{Precision-recall curve for best classifiers for test problems A and B.}
  \label{fig:curve}
\end{figure}

In all our experiments we used the same clustering algorithm, but, depending on the values of the hyperparameters, we obtained 96 different clustering patterns for each problem. We don't have a proper way to evaluate the quality of clusters themselves, so we compare the quality of the final classification. The validation dataset was used to compare the quality of approaches and find the best one. All in all, we evaluated 27648 combinations of different parameters. Then, best configurations were applied to the test dataset to get an independent assessment. Approaches based on the cosine similarity of the bag-of-words model and the autoencoder embeddings of the edit scripts demonstrated the best results on our dataset for problems A and B respectively. PR-curves for these classifiers are shown in Figure~\ref{fig:curve}.

\section{Conclusion}
In this paper, we present a method for the automatic classification of error types in solutions to programming assignments at an online learning platform. It is based on a notion of an edit script: we cluster these edit scripts at the training stage and classify newly submitted incorrect solutions according to these clusters. Manual labeling of these clusters allows us to provide users with a hint containing an error description. We provide an extensive evaluation of the proposed approach using various clustering methods, edit scripts representations and distance metrics. The evaluation shows that this approach could be successfully implemented in online programming courses at scale.

For the future work, we consider more advanced techniques for embedding atomic changes and edit scripts, e.g., RNN\cite{lipton2015critical}. It is also worthwhile to study the proposed method on a larger number of different problems and to identify characteristics of the dataset that could improve the quality of this approach.

\bibliographystyle{splncs04}
\bibliography{resources}
\end{document}